\documentclass[english,nocopyrightspace]{sigplanconf}
\usepackage{lmodern}

\usepackage[T1]{fontenc}
\usepackage[latin9]{inputenc}
\usepackage{babel}
\usepackage{verbatim}
\usepackage{varioref}
\usepackage{amsthm}
\usepackage{amsmath}
\usepackage{amssymb}
\usepackage{graphicx}
\usepackage[numbers]{natbib}

\makeatletter
\theoremstyle{plain}
\newtheorem{thm}{\protect\theoremname}
\theoremstyle{definition}
\newtheorem{defn}[thm]{\protect\definitionname}
\newcommand{\code}[1]{\texttt{#1}}

\@ifundefined{showcaptionsetup}{}{%
 \PassOptionsToPackage{caption=false}{subfig}}
\usepackage{subfig}
\makeatother

\usepackage{listings}
\providecommand{\definitionname}{Definition}
\providecommand{\theoremname}{Theorem}

\begin{document}

\title{Partial Cartesian Graph Product\\
{\Large{}(and its use in modeling Java subtyping)}}

\authorinfo{Moez A. AbdelGawad}
{Informatics Research Institute, SRTA-City, Alexandria, Egypt}
{\texttt{moez@cs.rice.edu}}
\maketitle
\begin{abstract}
\global\long\def\pcgp{\ltimes}
\global\long\def\cgp{\square}
\global\long\def\setcp{\times}
\global\long\def\setdu{+}
\global\long\def\cgu{\dotplus}
\global\long\def\adj{\sim}
In this paper we define a new product-like binary operation on directed
graphs, and we discuss some of its properties. We also briefly discuss
its application in constructing the subtyping relation in generic
nominally-typed object-oriented programming languages.
\end{abstract}


\section{Introduction}

Computer science is one of the many fields in which graph products
are becoming commonplace~\citep{Hammack2011}, where graph products
are often viewed as a convenient language with which to describe structures.
The notion of a product in any mathematical science enables the combination
or decomposition of its elemental structures. In graph theory there
are four main products: the Cartesian product, the direct/tensor/categorical
product, the strong product and the lexicographic product, each with
its own set of applications and theoretical interpretations.%

The applications of graph theory and graph products in researching
programming languages, in particular, are %
numerous%
\footnote{As revealed, for example, by doing an online search on `graph theory
and programming languages research'.}. In this paper we present a notion of a \emph{partial }Cartesian
graph product and discuss some of its properties.

We conjecture partial\emph{ }Cartesian graph products may have a number
of applications and uses in computer science, mathematics, and elsewhere.
In particular, we briefly demonstrate how the notion of a partial
Cartesian graph product we present in this paper can be applied to
accurately construct the subtyping relation in generic nominally-typed
object-oriented (OO) programming languages such as Java~\citep{JLS14},
C\#~\citep{CSharp2015}, C++~\citep{CPP2011}, Scala~\citep{Odersky14}
and Kotlin~\citep{Kotlin18}.

As such, this paper is structured as follows. In Section~\ref{sec:PCGP}
we present the definition of the partial Cartesian graph product of
two graphs and the intuition behind it (we present two equivalent
views of the partial product), then in Section~\ref{sec:Examples}
we present examples for partial Cartesian graph products that illustrate
our definition (in Appendix~\ref{sec:SageMath-Code} we present SageMath
code implementations of our definition/intuitions). In Section~\ref{sec:Basic-Properties}
we then discuss some of the basic properties of partial Cartesian
graph products.

The, in Section~\ref{sec:Related-Work}, we discuss some earlier
work similar to ours, and discuss the similarities and differences
between their properties. In Section~\ref{sec:Application} we then
discuss, in brief, how the partial Cartesian graph product operation
can be used to construct the subtyping relation in Java\footnote{We discuss this application in much more detail in~\citep{AbdelGawad2018b,AbdelGawad2018c}.}.
We conclude in Section~\ref{sec:Concluding-Remarks} with some final
remarks and a brief discussion of some research that can possibly
extend the theoretical and practical reach of the research presented
in this paper.

\section{\label{sec:PCGP}\label{sub:Definition}Partial Cartesian Graph Product}
\begin{defn}
\label{def:pcgp}(Partial Cartesian Graph Product, $\pcgp$). For
two %
directed graphs $G_{1}=(V_{1},E_{1})$ and $G_{2}=(V_{2},E_{2})$
where \end{defn}
\begin{itemize}
\item $V_{1}=V_{p}\setdu V_{n}$ such that $V_{p}$ and $V_{n}$ partition
$V_{1}$ (\emph{i.e.}, $V_{p}\subseteq V_{1}$ and $V_{n}=V_{1}\backslash V_{p}$),
\item $E_{1}=E_{pp}\setdu E_{pn}\setdu E_{np}\setdu E_{nn}$ such that $E_{pp}$,
$E_{pn}$, $E_{np}$, and $E_{nn}$ partition $E_{1}$,
\item $G_{p}=(V_{p},E_{pp})$ and $G_{n}=(V_{n},E_{nn})$ are two disjoint
subgraphs of $G_{1}$ (the ones induced by $V_{p}$ and $V_{n}$,
respectively, which guarantees that edges of $E_{pp}$ connect only
vertices of $V_{p}$ and edges of $E_{nn}$ connect only vertices
of $V_{n}$), and $E_{pn}$ and $E_{np}$ connect vertices from $V_{p}$
to $V_{n}$ and vice versa, respectively, and
\item $G_{2}$ is any directed graph (\emph{i.e.}, $G_{2}$, unlike $G_{1}$,
need not have some partitioning of its vertices and edges),
\end{itemize}
we define the\emph{ partial Cartesian graph product} of $G_{1}$ and
$G_{2}$ relative to the set of vertices $V_{p}\subseteq V_{1}$ as
\begin{equation}
G=G_{1}\pcgp_{V_{p}}G_{2}=(V,E)=G_{p}\cgp G_{2}\cgu G_{n}\label{eq:pcgp}
\end{equation}
 where
\begin{itemize}
\item $V=V_{p}\setcp V_{2}\setdu V_{n}$ ($\setcp$ and $\setdu$ are the
standard Cartesian set product and disjoint union operations),
\item $G_{p}\cgp G_{2}=(V_{p2},E_{p2})$ is the standard Cartesian graph
product~\citep{Hammack2011} of $G_{p}$ and $G_{2}$, and,
\item for defining $E$, the operator $\cgu$ is defined (implicitly relative
to $G_{1}$) such that we have\footnote{We may call $\cgu$ a ``Cartesian disjoint union'' (hence the addition-like
symbol $\cgu$), since $\cgu$ effects adding or attaching subgraph
$G_{n}$ to the Cartesian product $G_{p}\cgp G_{2}$, using edges
between $G_{n}$ and $G_{p}$ (in $G_{1}$) in the same way as these
edges are used to define edges in the Cartesian product $G_{1}\cgp G_{2}$.} 
\[
\begin{cases}
(u{}_{1},v_{1})\adj(u_{2},v_{2})\in E & \textrm{if }(u{}_{1},v_{1})\adj(u_{2},v_{2})\in E_{p2}\\
(u_{1},v)\adj u_{2}\in E & \textrm{if }u_{1}\adj u_{2}\in E_{pn},v\in V_{2}\\
u_{1}\adj(u_{2},v)\in E & \textrm{if }u_{1}\adj u_{2}\in E_{np},v\in V_{2}\\
u_{1}\adj u_{2}\in E & \textrm{if }u_{1}\adj u_{2}\in E_{nn}
\end{cases}
\]

\end{itemize}
Notes:
\begin{itemize}
\item As expressed by the definition of the partial Cartesian graph product,
each edge $e\in E$ in $G_{1}\pcgp_{V_{p}}G_{2}$ falls under exactly
one of four cases: either $e$ comes from $\ensuremath{G_{p}\cgp G_{2}}$,
or $e$ connects $\ensuremath{G_{p}\cgp G_{2}}$ to $\ensuremath{G_{n}}$,
or $e$ connects $\ensuremath{G_{n}}$ to $\ensuremath{G_{p}\cgp G_{2}}$,
or $e$ comes from $\ensuremath{G_{n}}$.
\item The vertices in set $V_{p}$ are called \emph{the} \emph{product vertices}
(of $G_{1}$), \emph{i.e.}, vertices that participate in the product
$G_{p}\cgp G_{2}$, while vertices in its complement, $V_{n}$ (which
we sometimes also write as $V'_{p}$), are called \emph{the non-product
vertices }(of $G_{1}$) since these vertices are not paired with vertices
of $G_{2}$ in the construction of $G_{1}\pcgp_{V_{p}}G_{2}$.
\item We call $G_{1}\pcgp_{V_{p}}G_{2}$ a \emph{partial} graph product
since, in comparison with the standard (full/total) Cartesian graph
product $G_{1}\cgp G_{2}$, the main component of $G_{1}\pcgp_{V_{p}}G_{2}$
(namely, the component $G_{p}\cgp G_{2}$) is typically the Cartesian
product of a \emph{proper }subgraph (namely, $G_{p}$) of $G_{1}$
with $G_{2}$.
\item Sometimes we omit the subscript $V_{p}$ and write $G_{1}\pcgp G_{2}$,
assuming $V_{p}$ is constant and implicit in the definition of $G_{1}$
(as is the case, for example, when using $\pcgp$ to model generic
OO subtyping).
\item In the partial graph product $G_{1}\pcgp_{V_{p}}G_{2}$, if we have
$V_{p}=V_{1}$ then we will have $G_{p}=G_{1}$ and $G_{n}$ will
be the empty graph, and in this case we have $G_{1}\pcgp_{V_{p}}G_{2}=G_{1}\cgp G_{2}$.
If, on the other hand, we have $V_{p}=\phi$ then $G_{p}$ will be
the empty graph and we will have $G_{n}=G_{1}$, and in this case
we have $G_{1}\pcgp_{V_{p}}G_{2}=G_{1}$.\\
In other words, in case all vertices of $G_{1}$ are product vertices
then, as might be expected, $G_{1}\pcgp_{V_{p}}G_{2}$ will be the
standard Cartesian product of $G_{1}$ and $G_{2}$, while in case
all vertices of $G_{1}$ are non-product vertices then $G_{1}\pcgp_{V_{p}}G_{2}$
will be just $G_{1}$ (\emph{i.e.}, graph $G_{2}$ is disregarded).
\end{itemize}

\paragraph*{\label{sub:Intuition}Intuition}

The intuition behind the definition of $\pcgp$ is simple. The partial
product $G_{1}\pcgp G_{2}$ of two graphs $G_{1}$ and $G_{2}$ can
be equivalently viewed as either:
\begin{itemize}
\item A graph that is based on the Cartesian product of the subgraph $G_{p}$
(of $G_{1}$) with $G_{2}$ that further includes $G_{n}$ while appropriately
respecting how $G_{n}$ is connected to $G_{p}$ in $G_{1}$ (which
is the view reflected in our definition of $\pcgp$ above\footnote{It is also the view reflected in our standard SageMath~\citep{Stein2017}
implementation of $\pcgp$. (See Appendix~\ref{sec:SageMath-Code}.)}), or as
\item Some sort of a special ``subgraph'' of the graph $G_{1}\cgp G_{2}$,
the standard Cartesian product of $G_{1}$ and $G_{2}$, where some
specified set of vertices of $G_{1}\cgp G_{2}$ (namely those of $V_{n}\times V_{2}$)
gets ``coalesced'' into a smaller set (one isomorphic to $V_{n}$),
\emph{i.e.}, where some vertices of $G_{1}$ (namely, vertices of
$G_{n}$, \emph{i.e.}, members of $V_{n}$) \emph{do not} \emph{fully}
\emph{participate} in the product graph (participate only with their
edges)\footnote{This was the view reflected in our initial SageMath implementation
of $\pcgp$. (See Appendix~\ref{sec:SageMath-Code}.)}.
\item The equivalence of these two informal views of $\pcgp$ can be proven
by showing that the product graphs resulting from the two views are
always isomorphic%
.
\end{itemize}

\section{\label{sec:Examples}Partial Graph Product Examples}

We illustrate the definition of $\pcgp$ by presenting the partial
Cartesian product of some sample graphs.

\begin{figure*}
\noindent \begin{centering}
\subfloat[$G$]{\noindent \protect\centering{}\protect\includegraphics[scale=0.4]{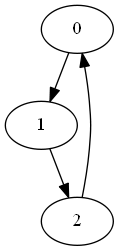}\protect}~~~~~~~~~~\subfloat[$G_{1}$]{\noindent \protect\centering{}\protect\includegraphics[scale=0.4]{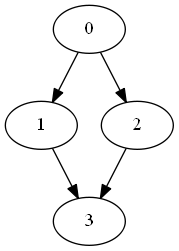}\protect}~~~~~~~~~~\subfloat[$G_{2}$]{\noindent \protect\centering{}\protect\includegraphics[scale=0.4]{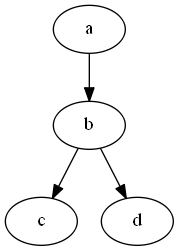}\protect}
\par\end{centering}

\protect\caption{\label{fig:Graphs-for-Illustrating}Graphs for illustrating $\protect\pcgp$}
\end{figure*}

\begin{figure}
\noindent \begin{centering}
\subfloat[$G\protect\pcgp_{\{\}}G=G$]{\noindent \protect\centering{}~~\protect\includegraphics[scale=0.4]{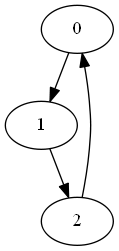}~~\protect}\subfloat[$G\protect\pcgp_{\{1\}}G_{1}$]{\noindent \protect\centering{}\protect\includegraphics[scale=0.3]{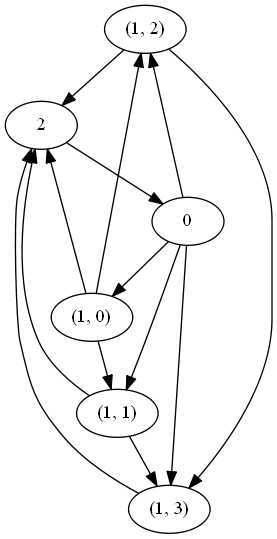}\protect}\subfloat[$G\protect\pcgp_{\{1\}}G$]{\noindent \protect\centering{}\protect\includegraphics[scale=0.3]{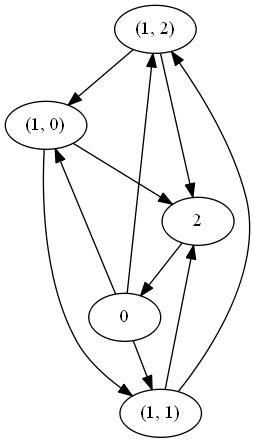}\protect}
\par\end{centering}

\noindent \begin{centering}
\subfloat[$G_{1}\protect\pcgp_{\{2,3\}}G_{2}$]{\noindent \protect\centering{}\protect\includegraphics[scale=0.3]{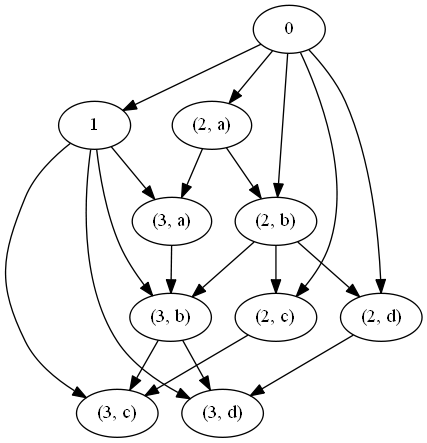}\protect}\subfloat[$G_{1}\protect\pcgp_{\{2,3\}}G$]{\noindent \protect\centering{}\protect\includegraphics[scale=0.3]{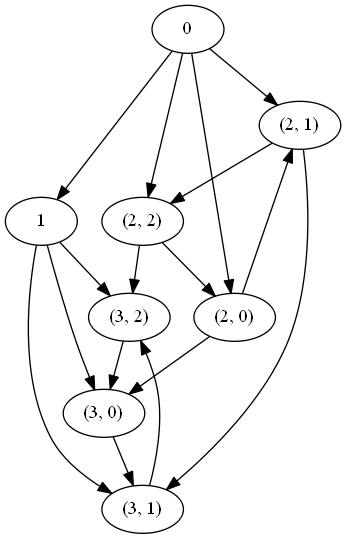}\protect}
\par\end{centering}

\noindent \centering{}\protect\caption{\label{fig:Partial-Product-Graphs}Partial product graphs (layout
by GraphViz)}
\end{figure}

\begin{figure}
\noindent \begin{centering}
\subfloat[$G_{1}\protect\pcgp_{\{2\}}G$]{\noindent \protect\centering{}\protect\includegraphics[scale=0.3]{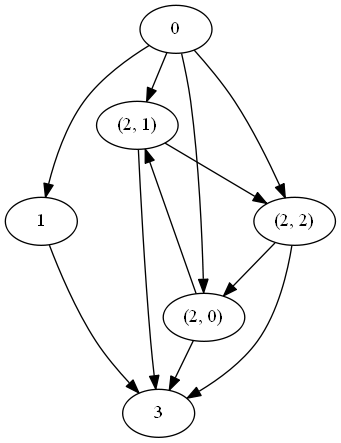}\protect}\subfloat[$G\protect\pcgp_{\{0,1,2\}}G=G\protect\cgp G$]{\noindent \protect\centering{}\protect\includegraphics[scale=0.25]{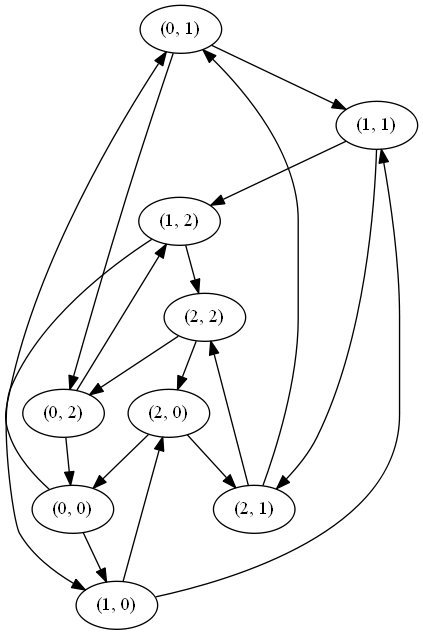}\protect}
\par\end{centering}

\noindent \centering{}\subfloat[$G\protect\pcgp_{\{2\}}G_{2}$]{\noindent \protect\centering{}\protect\includegraphics[scale=0.3]{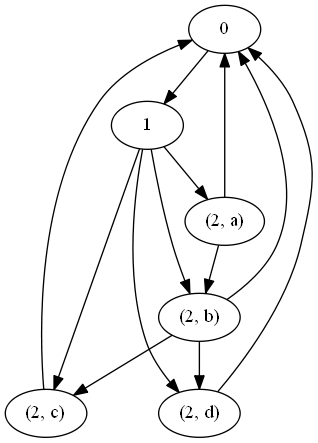}\protect}\subfloat[$G_{1}\protect\pcgp_{\{2\}}G_{2}$]{\noindent \protect\centering{}\protect\includegraphics[scale=0.3]{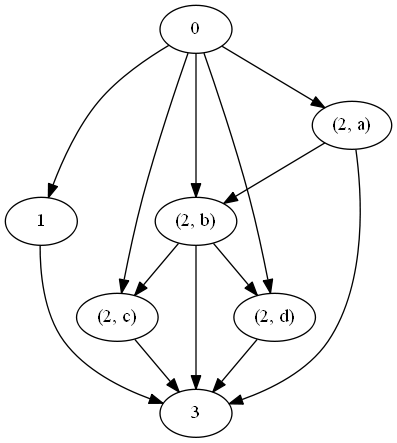}\protect}\protect\caption{\label{fig:Partial-Product-Graphs-1}Partial product graphs (layout
by GraphViz)}
\end{figure}

Consider the graphs $G$, $G_{1}$ and $G_{2}$ depicted in Figure~\ref{fig:Graphs-for-Illustrating}.
Figures~\ref{fig:Partial-Product-Graphs} and~\ref{fig:Partial-Product-Graphs-1}
present the graphs of some partial products of $G$, $G_{1}$ and
$G_{2}$. The reader should ensure he or she sees the product graphs
in Figures~\ref{fig:Partial-Product-Graphs} and~\ref{fig:Partial-Product-Graphs-1}
as intuitively clear\footnote{Even though some better layout of the graphs could make their task
even easier.}.

Appendix~\ref{sec:SageMath-Code} presents the SageMath code we used
to help generate the diagrams in Figures~\ref{fig:Partial-Product-Graphs}
and~\ref{fig:Partial-Product-Graphs-1}.

\section{\label{sec:Basic-Properties}Basic Properties of $\protect\pcgp$}

In this section we discuss some of the fundamental properties of partial
Cartesian graph products, particularly the size and order of constructed
graphs.

To calculate the number of vertices and number of edges in partial
product graphs, let $|S|$ denote the size (\emph{i.e.}, cardinality)
of a set $S$, and for a graph $G=(V,E)$ let $|G|=|V|$ denote the
number of vertices in $G$ (usually also called the size of $G$)
and let $\left\langle G\right\rangle =|E|$ denote the number of edges
(usually called the order of $G$).

Then, for a graph $G_{1}=(V_{1},E_{1})$ with size $v_{1}$ and order
$e_{1}$, a graph $G_{2}=(V_{2},E_{2})$ with size $v_{2}$ and order
$e_{2}$, and for a set $V_{p}\subseteq V_{1}$ with size $v_{p}\leq v_{1}$
and a complement $V'_{p}=V_{1}\backslash V_{p}$ with size $v'_{p}=v_{1}-v_{p}$
that induces a partitioning of $E_{1}=E_{pp}+E_{pn}+E_{np}+E_{nn}$
such that $G_{n}=(V'_{p},E_{nn})$, $e_{p}=|E_{pp}|+|E_{pn}|+|E_{np}|$
(as in Definition~\ref{def:pcgp}) and $e'_{p}=e_{1}-e_{p}=|E_{nn}|$
(\emph{i.e.}, $e_{1}=e_{p}+e'_{p}$), the number of vertices of the
partial Cartesian product graph is expressed by the equation
\begin{eqnarray}
|G_{1}\pcgp_{V_{p}}G_{2}|= & |V_{p}|\cdot|G_{2}|+|V'_{p}| & =v_{p}\cdot v_{2}+v'_{p}\label{eq:vertices}
\end{eqnarray}
while the number of edges is expressed by the equation
\begin{eqnarray}
\left\langle G_{1}\pcgp_{V_{p}}G_{2}\right\rangle  & = & |V_{p}|\cdot\left\langle G_{2}\right\rangle +\left(\left\langle G_{1}\right\rangle -\left\langle G_{n}\right\rangle \right)\cdot|G_{2}|+\left\langle G_{n}\right\rangle \nonumber \\
 & = & (v_{p}\cdot e_{2}+e_{p}\cdot v_{2})+e'_{p}\label{eq:edges}
\end{eqnarray}

Note that we also have 
\begin{eqnarray*}
\left\langle G_{1}\pcgp_{V_{p}}G_{2}\right\rangle  & = & (v_{p}\cdot e_{2}+e_{1}\cdot v_{2})-e'_{p}\cdot(v_{2}-1)\\
 & = & |V_{p}|\cdot\left\langle G_{2}\right\rangle +\left\langle G_{1}\right\rangle \cdot|G_{2}|-\left\langle G_{n}\right\rangle \cdot|G_{2}|+\left\langle G_{n}\right\rangle 
\end{eqnarray*}
which could be a more intuitive equation for $\left\langle G_{1}\pcgp_{V_{p}}G_{2}\right\rangle $
given that it indicates that edges of the partial product connecting
vertices of the product corresponding to $G_{n}$ get ``coalesced''
into one edge (\emph{i.e.}, multiedges are disallowed).%

For the sake of comparison, for the standard Cartesian product $G_{1}\cgp G_{2}$
(which \emph{is} a commutative operation, up to graph isomorphism)
we have 
\begin{eqnarray*}
|G_{1}\cgp G_{2}|= & |G_{1}|\times|G_{2}| & =v_{1}\cdot v_{2}\\
\left\langle G_{1}\cgp G_{2}\right\rangle = & |G_{1}|\times\left\langle G_{2}\right\rangle +\left\langle G_{1}\right\rangle \times|G_{2}| & =v_{1}\cdot e_{2}+e_{1}\cdot v_{2}.
\end{eqnarray*}

As we briefly illustrate in Section~\ref{sec:Application}, the fact
that the size of $G_{1}\pcgp_{S}G_{2}$ can be smaller than the multiplication
of the sizes of $G_{1}$ and $G_{2}$ (as in the standard Cartesian
graph product) makes $\pcgp$ perfectly suited for modeling generic
OO subtyping.

Note that, in the equations above, we intentionally depart from the
more common notation for graph sizes where $n$ is used to denote
the size of a graph and $m$ is used to denote its order, so as to
make the equations for sizes and orders of product graphs, particularly
Equations~(\ref{eq:vertices}) and~(\ref{eq:edges}), readily memorizable
and reminiscent of the graph equations defining the product graphs
themselves (\emph{e.g.}, Equation~(\ref{eq:pcgp})\vpageref{eq:pcgp}).

\section{\label{sec:Related-Work}Related Work}

The closest work to our work in this paper%
{} seems to be that of~\citep{Yero2015}. In~\citep{Yero2015} a definition
of another partial Cartesian graph product operation, denoted $\cgp_{S}$,
is presented.\footnote{We had not known this work existed until after we defined $\pcgp$
and named it.} Driven by our use of the partial Cartesian graph product $\pcgp$
in constructing the generic OO subtyping relation, our definition
of $\pcgp$ differs from that of $\cgp_{S}$ presented in~\citep{Yero2015},
as we present below.

\subsection{A Comparison of $\protect\pcgp$ and $\protect\cgp_{S}$}

First, it should be noted that the order of the factors $G_{1}$ and
$G_{2}$ in the partial products $G_{1}\pcgp_{S}G_{2}$ and $G_{2}\cgp_{S}G_{1}$
is reversed (due to the set $S$ being a subset of the vertices of
graph $G_{1}$, compared to graph $G_{2}$ graph $G_{1}$ has a special
status in the products, and thus both partial products are non-commutative
operations. For both operations, the order of the factors of the products
matters).

More significantly, as we explain using equations in the sequel, while
$G_{1}\pcgp_{S}G_{2}$ and $G_{2}\cgp_{S}G_{1}$ can have the same
number of edges, $G_{1}\pcgp_{S}G_{2}$ typically has less vertices
than $G_{2}\cgp_{S}G_{1}$.

Using the same notation as that of Section~\ref{sec:Basic-Properties},
the number of vertices of a partial product graph $G_{2}\cgp_{V_{p}}G_{1}$
is expressed by the equation
\begin{eqnarray*}
|G_{2}\cgp_{V_{p}}G_{1}|= & |G_{2}|\cdot|G_{1}| & =v_{2}\cdot v_{1}
\end{eqnarray*}
while the number of its edges is expressed by the equation
\begin{eqnarray*}
\left\langle G_{2}\cgp_{V_{p}}G_{1}\right\rangle  & = & |G_{2}|\cdot\left\langle G_{1}\right\rangle +\left\langle G_{2}\right\rangle \cdot|V_{p}|\\
 & = & v_{2}\cdot e_{1}+e_{2}\cdot v_{p}\\
 & = & v_{p}\cdot e_{2}+e_{1}\cdot v_{2}.
\end{eqnarray*}

Note also that if multiedges were allowed for $\pcgp$ we would have
\[
\left\langle G_{1}\pcgp_{V_{p}}G_{2}\right\rangle =v_{p}\cdot e_{2}+e_{p}\cdot v_{2}
\]
and the order of $G_{1}\pcgp_{V_{p}}G_{2}$ will then be the same
as that of $G_{2}\cgp_{V_{p}}G_{1}$ (which, when multiedges are disallowed,
happens only if $G_{n}$ has no edges, \emph{i.e.}, when none of the
vertices of $G_{n}$ is connected to another vertex of $G_{n}$, sometimes
called a discrete graph).

Also it should be noted that the full (\emph{i.e.}, standard) Cartesian
graph product can be obtained using either of the two partial Cartesian
graph products by setting $V_{p}=V_{1}$. This illustrates that, compared
to the standard Cartesian graph product, if $V_{p}\neq V_{1}$ then
the partial product operation $\pcgp$ decreases both the vertices
and the edges of the product while the partial product operation $\cgp_{V_{p}}$
decreases only the edges of the product.

\begin{figure}
\noindent \begin{centering}
\subfloat[$G$]{\noindent \protect\centering{}\protect\includegraphics[scale=0.5]{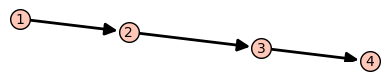}\protect}
\par\end{centering}

\noindent \begin{centering}
\subfloat[$G\protect\pcgp_{\{1,3\}}G$]{\noindent \protect\centering{}\protect\includegraphics[scale=0.5]{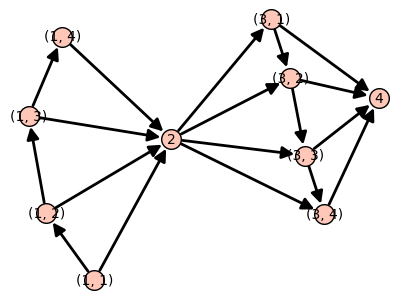}\protect}
\par\end{centering}

\noindent \begin{centering}
\subfloat[$G\protect\cgp_{\{1,3\}}G$]{\noindent \protect\centering{}\protect\includegraphics[scale=0.5]{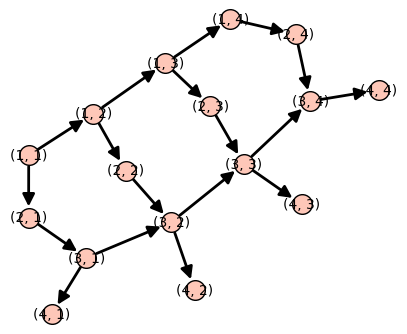}\protect}
\par\end{centering}

\noindent \begin{centering}
\subfloat[$G\protect\pcgp_{\{1,4\}}G$]{\noindent \protect\centering{}\protect\includegraphics[scale=0.5]{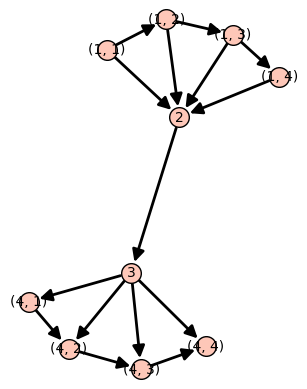}\protect}
\par\end{centering}

\noindent \begin{centering}
\subfloat[$G\protect\cgp_{\{1,4\}}G$]{\noindent \protect\centering{}\protect\includegraphics[scale=0.5]{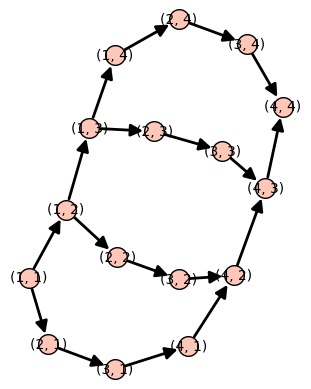}\protect}
\par\end{centering}

\protect\caption{\label{fig:Comparing}Comparing $\protect\pcgp_{S}$ to $\protect\cgp_{S}$
(layout by SageMath)}
\end{figure}

To visually illustrate the difference between $\pcgp$ and $\cgp_{S}$
we adapt the example presented in~\citep{Yero2015} for illustrating
$\cgp_{S}$. The graph diagrams presented in Figure~\ref{fig:Comparing}
help illustrate the differences between the two operations we discussed
above.

Further adding to the differences between $\pcgp$ and $\cgp_{S}$,
the main motivation for defining $\pcgp$ is to apply it in modeling
generic OO subtyping, while the motivation behind defining $\cgp_{S}$---
as presented in~\citep{Yero2015}---seems to be a purely theoretical
motivation, namely, studying Vizing's conjecture (a famous conjecture
in graph theory, relating the domination number of a product graph
to the domination number of its factors).

Finally, our choice of the symbol $\pcgp$ for denoting the partial
product operation allows for making $S$ implicit while indicating
that the product operation is partial. For the notation $\cgp_{S}$
doing this is not possible, given that the symbol $\cgp$---which
will result if $S$ is dropped from the notation---is the symbol for
the standard Cartesian graph product, \emph{i.e.}, for a different
operation.

\section{\label{sec:Application}An Application of $\protect\pcgp$: Modeling
Generic OO Subtyping}

Generic types~\citep{JLS05,JLS14,GenericsFAQWebsite,AbdelGawad2016a,AbdelGawad2016c,AbdelGawad2017b}
add to the expressiveness and type safety of industrial-strength object-oriented
programming (OOP) languages such as Java, C\#, Scala, Kotlin and other
nominally-typed OO programming languages~\citep{AbdelGawad2015}.
As we detail in~\citep{AbdelGawad2016c,AbdelGawad2017b}, many models
for generics have been proposed, particularly for modeling features
such as wildcard types~\citep{Torgersen2004,MadsTorgersen2005,Cameron2007,Cameron2008,Summers2010,Tate2011,Tate2013,Greenman2014}.
However, as expressed by their authors, none of these models seem
to be a fully satisfactory model.

This situation, in our opinion, is due to these models and the mathematical
foundations they build upon distancing themselves (unnecessarily)
from the nominal-typing of generic OOP languages and, accordingly,
them being unware of the far-reaching implications nominal-typing
has on the type systems of these languages and on analyzing and understanding
them, which---again, in our opinion---includes analyzing and understanding
generics and generic variance annotations (of which wildcard types
are instances).

To demonstrate the direct effect of nominal-typing on the Java type
system and on generics in particular, we illustrate how the generic
\emph{subtyping} relation in Java can be constructed, using $\pcgp$
and the \emph{subclassing} relation (which is an inherently nominal
relation, in Java and in all OO languages) based on the nominality
of the subtyping relation in Java (\emph{i.e.}, due to the nominal
typing and nominal subtyping of Java, the subclassing relation is
the basis for defining the subtyping relation).

In brief, with some simplifying assumptions that we detail in~\citep{AbdelGawad2017a,AbdelGawad2018b},
the generic subtyping relation in Java can be constructed iteratively
using the nominal subclassing relation and the partial Cartesian graph
product $\pcgp$, as follows.

Let $C$ be the graph of the subclassing relation in some Java program.
Let $C_{g}$ be the generic classes subset of classes in$C$. Then
the graph $S$ of the subtyping relation in the Java program (typically
$S$ is infinite, if there is at least one generic class in $C$)
can be constructed as the limit of the sequence of graphs $S_{i}$
of subtyping relations constructed iteratively using the equation
\begin{equation}
S_{i+1}=C\pcgp_{C_{g}}S_{i}^{\triangle}\label{eq:gs-pcgp}
\end{equation}
where $S_{i}^{\triangle}$ is the graph of the containment relation
between wildcard type arguments derived from $S_{i}$ (as explained
in~\citep{AbdelGawad2018b}) and $S_{0}^{\triangle}=Graph(`\mathtt{?}\textrm{'})$
is the one-vertex graph having the default wildcard type argument,
`\code{?}', as its only vertex and no containment relation edges
(again as explained in~\citep{AbdelGawad2018b}).

It should be noted that Equation~(\ref{eq:gs-pcgp}) tells us that
in the construction of the graph of the subtyping relation $S$ \emph{the
generic classes in $C$} (\emph{i.e.}, $C_{g}$)\emph{ correspond
to product vertices}, while \emph{the non-generic classes in $C$
correspond to non-product vertices} in the partial product graph of
each approximation $S_{i+1}$ of $S$.\footnote{This observation has been a main motivation behind our definition
of $\pcgp$.} This property of $\pcgp$ preserves non-generic types (and the subtyping
relations between them) during the construction of $S$, meaning that
non-generic types in $S_{i}$ remain as non-generic types in $S_{i+1}$,
and thus, ultimately, are non-generic types in $S$ as well.

\subsection{Java Subtyping Example}

\begin{figure}
\noindent \begin{centering}
\subfloat[$C$]{\noindent \protect\centering{}\protect\includegraphics[scale=0.5]{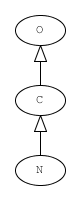}\protect}~~~~~\subfloat[$S_{1}=C\protect\pcgp_{\{\mathtt{C}\}}S_{0}^{\triangle}$]{\noindent \protect\centering{}~~~~~\protect\includegraphics[scale=0.5]{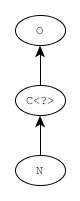}~~~~~\protect}~~~~~\subfloat[$S_{2}=C\protect\pcgp_{\{\mathtt{C}\}}S_{1}^{\triangle}$]{\noindent \protect\centering{}\protect\includegraphics[scale=0.3]{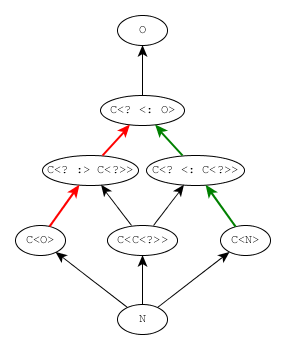}\protect}
\par\end{centering}

\protect\caption{\label{fig:Illustrating-the-use}A simple illustration of the use
of $\protect\pcgp$ to model generic OO subtyping (manual layout using
yEd)}
\end{figure}

Figure~\ref{fig:Illustrating-the-use} illustrates the use of $\pcgp$
to construct the Java subtyping relation. To decrease clutter, given
that OO subtyping is a transitive relation, we present the transitive
reduction of the subtyping graphs in Figure~\ref{fig:Illustrating-the-use}.

The three graphs in Figure~\ref{fig:Illustrating-the-use} illustrate
the construction of the subtyping relation $S$ of a Java program
that only has the generic class definition

\code{\textbf{class}~C<T>~\{\}}

As defined by Equation~(\ref{eq:gs-pcgp}), the graph of $S_{2}=C\pcgp_{\{\mathtt{C}\}}S_{1}^{\triangle}$
in Figure~\ref{fig:Illustrating-the-use} is constructed as the partial
product of the graph of the subclassing/inheritance relation $C$
and the graph of $S_{1}^{\triangle}$ (of wildcard types over $S_{1}$,
ordered by containment) relative to the set $\{\mathtt{C}\}$ of generic
classes in $C$.

More details and examples on the use of $\pcgp$ to construct the
generic OO subtyping relation can be found in~\citep{AbdelGawad2018b,AbdelGawad2018c}.

\section{\label{sec:Concluding-Remarks}Concluding Remarks and Future Work}

In this paper we defined a new binary operation $\pcgp$ on graphs
that constructs a partial product of its two input graphs, we presented
few examples that illustrate the definition of $\pcgp$, and we discussed
some of the basic properties of the operation. We also compared the
$\pcgp$ operation to the closest similar work. Finally, we also discussed
how the partial graph product operation $\pcgp$ may be used in understanding
the subtyping relation in generic nominally-typed OO programming languages.

As of the time of this writing, we do not know of any other application
of the new graph operation we present. Nevertheless, in this paper
we presented the partial product operation over graphs in abstract
mathematical terms, in the hope that the operation may prove to be
useful in other mathematical contexts and domains.

Although we have not done so here, we believe the notion of partial
Cartesian graph products, as presented here, can be easily adapted
to apply to other mathematical notions such as sets, partial orders,
groups (or even categories, more generally). To model infinite self-similar
graphs (or groups or categories) we also believe partial products,
over graphs, groups, or categories, can in some way be modeled by
operads, which are category-theoretic tools that have proved to be
useful in modeling self-similar phenomena~\citep{spivak2014category,AbdelGawad2017a}.

Finally, studying in more depth properties of partial Cartesian graph
products such as the size, order (as we hinted at in Section~\ref{sec:Related-Work})
and rank of elements of the products and of infinite applications
of them, is work that can build on work we presented in this paper,
and which can be of both theoretical and practical significance, particularly
in computer science graph theoretic applications. Also, we believe
a notion of `degree of partialness' of a partial product graph\footnote{For example, the degree of partialness can be a value (a real number)
between 0 and 1, defined possibly as the size of the set of product
vertices, $|V_{p}|$, divided by size of the set of all vertices of
the first factor graph, $|V_{1}|$ (\emph{i.e.}, the degree of partialness
of a partial product will be $|V_{p}|/|V_{1}|$. In this case a degree
of partialness with value 1 means the standard Cartesian product,
while a value of 0 means no product.)} can be a useful notion, even though we do not immediately see an
application of this notion.

\bibliographystyle{plain}

\appendix

\section{\label{sec:SageMath-Code}SageMath Code}

To generate the graph examples presented in this paper we implemented
the definition of $\pcgp$ (as presented in Section~\ref{sub:Definition})
in SageMath 8.1~\citep{Stein2017}. For those interested, we present
in this appendix our SageMath implementation code. The code presented
here is not optimized for speed of execution but rather for clarity
and simplicity of implementation.

\begin{lstlisting}[language=Python,basicstyle={\small\ttfamily},frame=lines]
# PCGP

def comp(pv,g):
  """ Computes the complement of pv relative to
      vertices of g
  """
  return filter(lambda v: v not in pv, 
                g.vertices())

def PCGP(g1,pv,g2):
  """ Computes the partial cartesian product of 
      graphs g1 and g2.

  INPUT:      
  
  - ``pv'' (list) -- is the list of product
    vertices in g1.   

  """
  # 1st step
  gp = g1.subgraph(pv)
  g = gp.cartesian_product(g2)

  # 2nd step
  npv = comp(pv,g1)
  gn = g1.subgraph(npv)
  g = g.union(gn)

  # 3rd step
  gpn = g1.subgraph(edge_property=
            (lambda e: e[0] in pv and e[1] in npv))
  for u1,u2 in gpn.edge_iterator(labels=None):
    for v in g2:
      g.add_edge((u1,v),u2)

  # 4th step
  gnp = g1.subgraph(edge_property=
            (lambda e: e[1] in pv and e[0] in npv))
  for u1,u2 in gnp.edge_iterator(labels=None):
    for v in g2:
      g.add_edge(u1,(u2,v))

  return g 
\end{lstlisting}

For convenience, our initial shorter (but equivalent) implementation
code for $\pcgp$ (where \code{GSP} stands for `Generic Subtyping
Product') is as follows. The code corresponds to the second informal
view of the partial Cartesian graph product we presented in Section~\ref{sub:Intuition}.

\begin{lstlisting}[language=Python,basicstyle={\small\ttfamily},frame=lines]
# GSP
def GSP(g1, pv, g2):
  g=DiGraph.cartesian_product(g1,g2) # main step

  lnpvc = map(lambda npv: filter(lambda(v,_): 
              v==npv, g.vertices()), comp(pv,g1))
  # lnpvc is list of non-product vertex clusters

  # merge the clusters
  map(lambda vc: g.merge_vertices(vc), lnpvc)

  return g
\end{lstlisting}

For any two graphs \code{g1}, \code{g2} and any list \code{pv}
(listing the product vertices subset of the vertices of \code{g1})
we have
\begin{lstlisting}[language=Python,basicstyle={\small\ttfamily},frame=lines]
  GSP(g1,pv,g2).is_isomorphic(PCGP(g1,pv,g2))
\end{lstlisting}

\end{document}